\documentclass[
    ,final            % use final for the camera ready runs
  ,numberedheadings % uncomment this option for numbered sections
  ] {aipproc}
\layoutstyle{8x11double}
\usepackage{amssymb, amsmath, amsbsy}
\usepackage[T1]{fontenc}
\usepackage{listings}
\usepackage{graphicx}
\usepackage{color}

\begin{document}

\title{How to Calculate Tortuosity Easily?}

\classification{47.56.$+$r, 47.15.G$-$, 91.60.Np}
\keywords      {tortuosity; porous media}

\author{Maciej Matyka}{
  address={Institute of Theoretical Physics, University of Wroc{\l}aw, pl. M. Borna 9,
           50-204 Wroc{\l}aw, Poland, tel.: +48713759357, fax: +48713217682}
}

\author{Zbigniew Koza}{
  address={Institute of Theoretical Physics, University of Wroc{\l}aw, pl. M. Borna 9,
           50-204 Wroc{\l}aw, Poland, tel.: +48713759357, fax: +48713217682}
}

\begin{abstract}
Tortuosity is one of the key
parameters describing the geometry and transport
properties
of porous media. It is defined either as an average elongation of fluid paths or
as a retardation factor that measures the resistance of a porous medium to the flow.
However, in contrast to a retardation factor, an average fluid path elongation is difficult to
compute numerically and, in general, is not measurable directly in experiments.
We review some recent achievements in bridging the gap between the two formulations of tortuosity and
discuss possible method of numerical and an experimental measurements of the tortuosity
directly from the fluid velocity field.
\end{abstract}

\maketitle
\section{Introduction}

One of the main problems in porous media physics is to find out how the value of permeability,
which synthetically describes flow retardation by the porous medium structure,
is correlated with
the geometry of the medium.
Another problem is to define macroscopic parameters that
could be used to distinguish various kinds of porous media.
One of the parameters that is often used for both of these purposes is the tortuosity.

The notion of tortuosity was introduced to porous media studies by Carman \cite{Carman37},
who considered a flow through a bed of sand and proposed
the tortuosity as a factor that accounts for effective elongation of fluid paths.
Assuming that
a porous bed of thickness $L$ can be regarded as a bundle of capillaries
of equal length $L_\mathrm{eff}$ and constant cross-section,
he proposed the following semi-empirical Kozeny-Carman formula \cite{Carman37,Bear72,Dullien92}
\begin{equation}
  \label{eq:Kozeny-Carman}
    k =  \frac{\varphi ^{3} }{\beta T^{2} S^{2} },
\end{equation}
which relates the permeability ($k$) to four structural parameters:
the porosity $\varphi$, the specific surface area $S$,
the shape factor $\beta$,
and the hydraulic tortuosity $T$,
\begin{equation}
  \label{eq:tortuosity}
    T = \frac{L_\mathrm{eff}}{L}.
\end{equation}

The simple capillary model used by Kozeny and Carman can be easily
applied to other forms of transport through porous media. For example, the electric tortuosity ($T_\mathrm{el}$)
is defined as a retardation factor \cite{Clennell97},
\begin{equation}
  \label{eq:tortuosity-el}
  T_\mathrm{el} = \frac{\sigma_\mathrm{fl}}{\sigma_\mathrm{p}},
\end{equation}
where $\sigma_\mathrm{fl}$ is the electrical conductivity of a conductive fluid
and $\sigma_\mathrm{p}$ is the effective electrical conductivities of a porous medium filled with this fluid.
Within the simple capillary model, $T_\mathrm{el}$ is related to $L_\mathrm{eff}/L$ through
\begin{equation}
  \label{eq:tort-electrical}
  T_\mathrm{el} = \frac{1}{\varphi}\left(\frac{L_\mathrm{eff}}{L}\right)^2,
\end{equation}
and a similar relation holds for the diffusive tortuosity \cite{Clennell97}.

\begin{figure}
\centering
\includegraphics[width=0.45\textwidth]{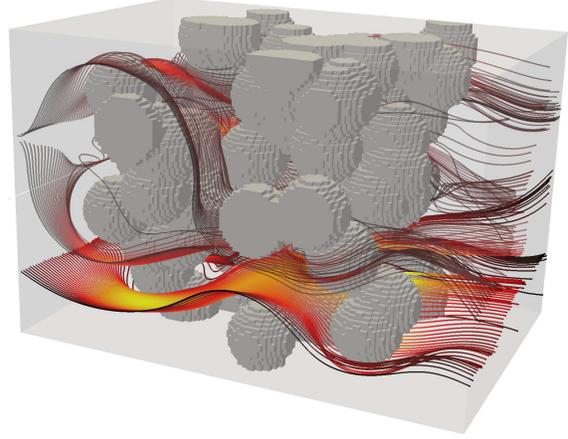}
\caption{Streamlines in the fluid flow through three-dimensional
random model of porous media at porosity $\varphi=0.6$ and tortuosity
T=1.15.\label{fig1}}
\end{figure}

Comparison of Eqs. (\ref{eq:tortuosity}) and (\ref{eq:tort-electrical}), as well as a research into the literature,
reveal the first problem with tortuosity: depending on the context,
this term can be related to $T$, $T^2$ or even $T^{-1}$ or $T^{-2}$
\cite{Bear72,Dullien92,Clennell97,Araujo06}.
The second problem is that a link between the tortuosity defined as an average elongation of fluid paths,
as in Eq. (\ref{eq:tortuosity}) (see Fig.~\ref{fig1}), and the tortuosity defined as a retardation factor,
as in Eq. (\ref{eq:tort-electrical}),  is well-defined only for the capillary model.
It is not clear that a similar correlation exists for arbitrary porous media.
The third problem is an imprecise definition of the effective fluid path length ($L_\mathrm{eff}$)
in Eq.~(\ref{eq:tortuosity}). In real porous media flow paths are extremely complicated,
as the fluid fluxes continuously change in sectional area, shape and orientation as well as branch and rejoin,
and this observation led several researchers \cite{Dullien92} to believe that $L_\mathrm{eff}$
can be defined only in relatively simple network models, which, however, can be analyzed without this notion.
The fourth problem is that the Carman-Kozeny formula, Eq.~(\ref{eq:Kozeny-Carman}), actually defines
the product $\beta T^2$ (known as the Kozeny constant) and if the tortuosity could not be
 measured for general porous media, the shape factor and hydraulic tortuosity
would become essentially indeterminate quantities, rendering the tortuosity a `fudge factor'
used to fit the model to the experimental data \cite{Dullien92,Tye82}.

In this paper we discuss some recent achievements in solving the above-mentioned problems.
In particular, we show some applications and implications of our recent formula for the tortuosity \cite{Duda11}
\begin{equation}
  \label{tformula}
  T=\frac{\langle u\rangle}{\langle u_x\rangle},
\end{equation}
where $\langle u \rangle$ is the average magnitude of the intrinsic velocity
over the entire system volume and $\left<u_x\right>$ is the volumetric average of its
component along the macroscopic flow direction.

\section{Solution to the fluid flow problem}

In order to compute the tortuosity defined with Eq.~(\ref{tformula}) one has to know the steady state velocity field. This
may be accomplished either experimentally, e.g.\ by using the particle
image velocimetry methods \cite{Cassidy05,Lachnab08, Morad09},
the magnetic resonanse imaging \cite{Irwin99,Werth10} or numerically, by finding the solution to
the Navier--Stokes equations in the pore space of a porous medium.

Here we take the numerical approach.
We use the lattice Boltzmann method (LBM) \cite{Succi01}. In this  method, which originates from the kinetic theory of gases and
the lattice gas automata models  \cite{McNamara88},
the fluid is modeled as consisting of fictive particles propagating and colliding on a discrete lattice.
Space, time and velocities are all discrete, with space usually discretized into a regular grid,
time discretized into equal intervals, and velocities restricted to just a few vectors $\mathbf{c}_i$
related  to the geometry of the lattice.
The state of the system is fully characterized by distribution functions
$f_i(\mathbf{x},t) \in [0,1]$ defined for each lattice node
$\mathbf{x}$, discrete time $t$, and $\mathbf{c}_i$.
They can be interpreted as being proportional to the number of
particles that at time $t$ are at node $\mathbf{x}$ and have velocity
$\mathbf{c}_i$.
It was shown that solving the LBM model leads to the solution of
the incompressible Navier-Stokes equations \cite{Luo00}.

\subsection{Sailfish and tortuosity computation}

All our numerical computations were performed using the Sailfish library \cite{website:sailfish}, which is an implementation of the LBM method
running on graphics processing units (GPUs), an emerging platform for high-performance computing. Sailfish is an open-source project written in the Python programming language, and hence is a highly customisable software.
In particular, implementation of Eq.\ (\ref{tformula}) in Sailfish is trivial and %its Sailfish variant implemented as a method of the main simulation class
consists of just a few lines of code.

\section{Tortuosity computation}

A general discussion of Eq.~(\ref{tformula}), its relation to Eq. (\ref{eq:tortuosity}) and conditions of applicability can be found in
\cite{Duda11}. Here we use the fact that the LBM method uses a regular (i.e. cubic) mesh. This allows to approximate  Eq.~(\ref{tformula}) with
\begin{equation}
  \label{tformula-sums}
  T=\frac{\sum_\mathbf{r} u(\mathbf{r})}{\sum_\mathbf{r} u_x(\mathbf{r})},
\end{equation}
where $\mathbf{r}$ runs through all lattice nodes. Note that this simple formula can be used not only in numerical studies, 
but also for the data obtained experimentally.

In subsequent subsections we test how Eq.\ (\ref{tformula-sums}) behaves for flows in various geometries of increasing complexity.

\subsection{Inclined channel}

The simplest model of a porous medium approximates it as
a bunch of straight pipes, each inclined at an angle $\alpha$
to the macroscopic flow direction. In this case all streamlines are of the same length and Eq.\ (\ref{eq:tortuosity}) yields $T= 1/\cos\alpha$.
\begin{figure}[!t]
\centering
\includegraphics[width=.348\textwidth]{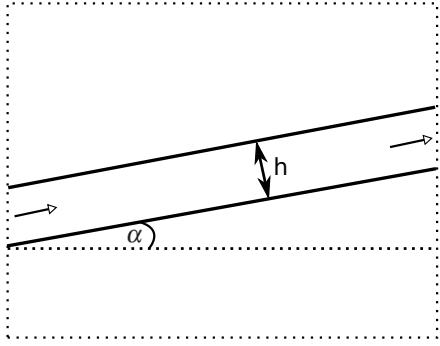}
\caption{A channel rotated by an angle $\alpha$.\label{inclinedscheme}}
\end{figure}
We constructed two two-dimensional (2D) configurations of an inclined channel
of height h, with $h=5$ and $20$ lattice units (l.u.) (see Fig.\
\ref{inclinedscheme}). We used the mesh
resolution $240 \times 800$ (l.u.), set the lattice kinematic viscosity at
$\nu=0.01$ and assumed the lattice velocity $u_{\mathrm{max}}=0.02$ as the maximum value of the developed velocity profile at the inlet and outlet. Starting from $\alpha=0$, we succesively rotated the channel preserving its width. For each inclination angle the steady state was found using the LBM method and then the tortuosity was calculated from  Eq.\ (\ref{tformula-sums}). The results are depicted in Fig.\ \ref{inclinedchannel}.
\begin{figure}[!t]
\centering
\includegraphics[width=.47\textwidth]{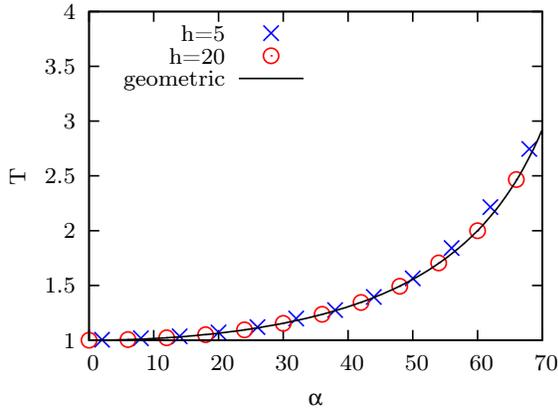}
\caption{Hydraulic tortuosity $T$, Eq.\ (\protect\ref{tformula}), in
a channel of height $h$ (lattice units) rotated by an angle $\alpha$.
The solid line shows the geometric tortuosity $1/\cos\alpha$.
\label{inclinedchannel}}
\end{figure}
They agree well with the values expected from geometric considerations even for a narrow channel of height h=$5$ l.u.
(the relative difference 
does not exceed 5\%).
Small errors observed in our simulations stem from
discretization errors and the staircased approximation of the channel boundaries in the LBM method \cite{Matyka12}, which results in $T$ being slightly overestimated. These errors decrease with an increasing value of $h$.

\subsection{U-shaped channel}

\begin{figure}
\centering
\includegraphics[width=.38\textwidth]{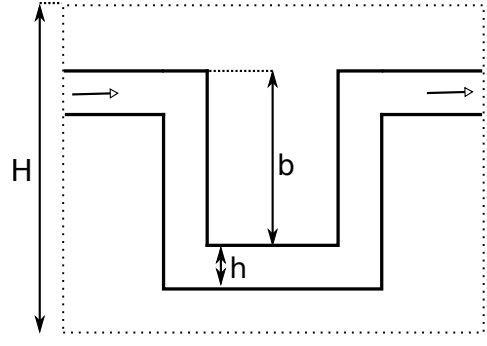}
\caption{The U-shaped channel with a step depth $b$.
\label{uchannelscheme}}
\end{figure}
Next, we constructed a U-shaped channel geometry (see
Fig.\ \ref{uchannelscheme}). The mesh resolution was set
at $L\times H = 300\times 200$ (l.u.), and we assumed $\nu=0.01$ and $u_{\mathrm{max}}=0.01$. We started from the step
depth $b=0$ and increased it until $b=H/2$.
For each  $b$ two channel heights, $h=5$ and
$h=20$ (l.u.) were investigated.
In this geometry 
the geometric tortuosity $T_\mathrm{g}=(L+2b)/L$.
\begin{figure}
\centering
\includegraphics[width=.47\textwidth]{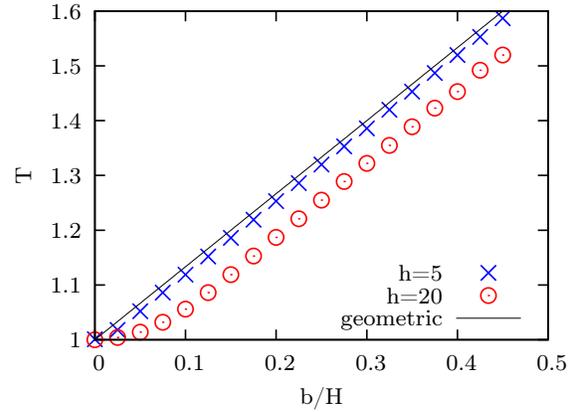}
\caption{Hydraulic tortuosity, $T$, computed using Eq.\ (\ref{tformula}) in
a U-shaped channel as a function of the relative step depth $b/H$ for two channel heights $h$ (lattice units).
The solid line represents the geometric tortuosity for the same
system.\label{uchannel}}
\end{figure}
%c
Comparison of the tortuosity determined from Eq.\ (\ref{tformula-sums})
with $T_\mathrm{g}$ for various values of $b$ and $h$
is shown in Fig.\ \ref{uchannel}.
As expected, a linear dependency of geometric and hydraulic tortuosities on $b$ is visible, but a deviation of $T$ from $T_\mathrm{g}$ at larger channel depths $b$ is also noticeable.
To understand this effect we analyzed the flow
streamlines for this systems (data not shown) and
observed that they follow the geometry nicely only at the
straight parts of the channel. At corners, however,
a tendency of flow paths to seek a shorter path is visible. This leads to
the hydraulic tortuosity $T$ being smaller than
the geometric one, as shown in Fig.\ \ref{uchannel}.
This effect is more visible for larger $h$, as in this case the corner deformation is more pronounced.

\subsection{Two-dimensional overlapping circles}

\begin{figure}
\centering
\includegraphics[width=.47\textwidth]{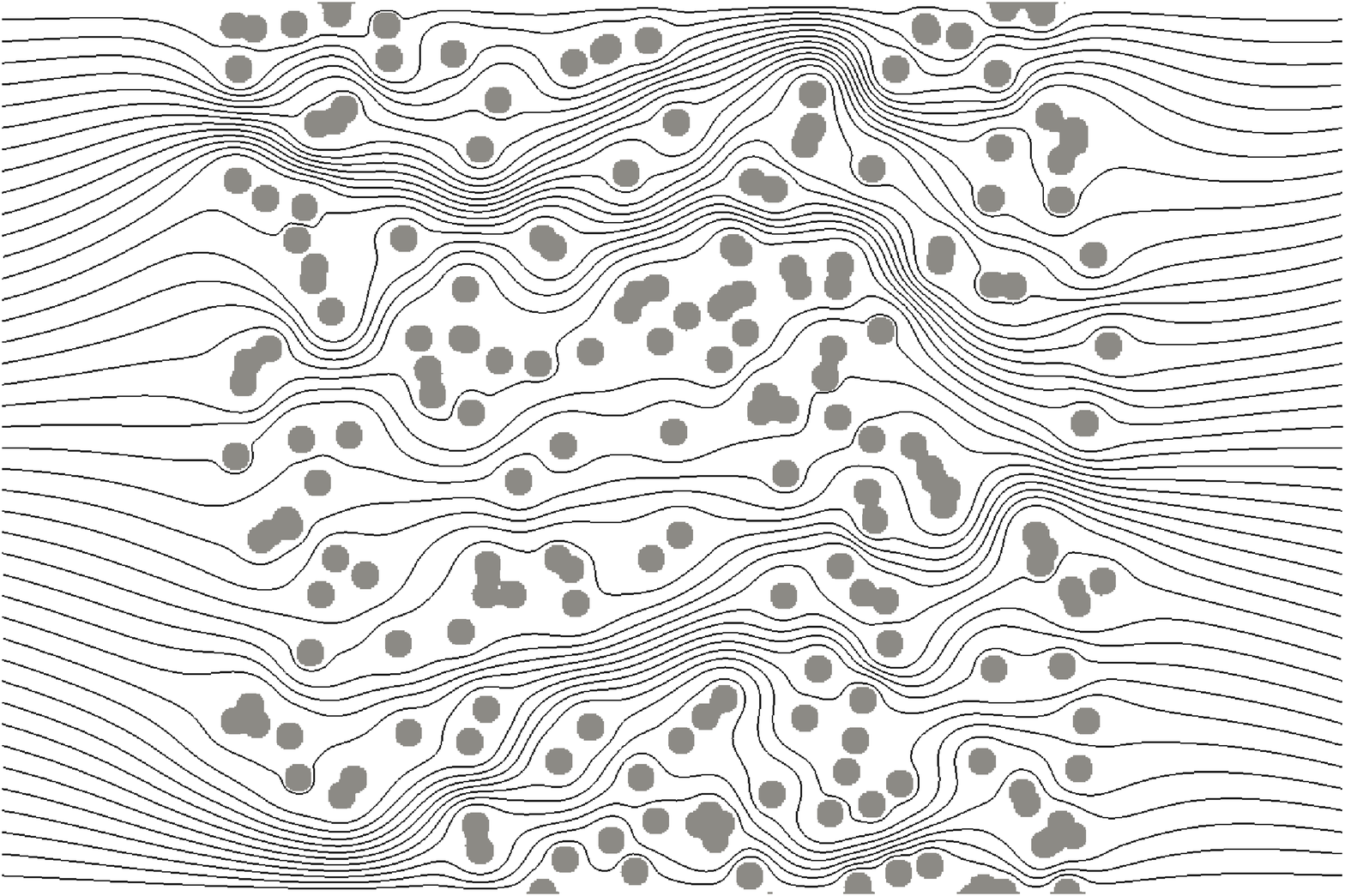}
\caption{Flow streamlines in a realization of a two-dimensional model of a porous
medium built of overlapping circles of radius $R=10$ (l.u.) at porosity
$\varphi=0.85$. The tortuosity computed using Eq.\ (\protect\ref{tformula-sums}) is
$T=1.13$. \label{tort2d}}
\end{figure}

The next model we considered was a 2D system in which the porous matrix was modeled by
circles that were free to overlap. We used a $900\times 600$ mesh at which we randomly
deposited monosized circles of radius $r=10$ l.u.
No-slip boundary conditions were imposed on the top and bottom walls and periodic
boundary conditions were assumed at the inlet and outlet. The flow was
driven by an external force field whose magnitude was chosen so that the Reynolds number
$\mathrm{Re}<1$ (creeping flow). The circles were deposited only in the central, $600\times 600$ (l.u.) area
of the mesh,  see  Fig.\ \ref{tort2d}.
The remaining space was kept empty
to minimize the influence of inlet and outlet boundary conditions. We
ran the simulation for 40~000 steps until the steady state was reached
and used Eq.\ (\ref{tformula-sums}) to compute the tortuosity.
As seen in Fig.\ \ref{tort2d}, the flow streamlines in this model can be quite complex
and ``tortuous''.

\begin{figure}
\centering
\includegraphics[width=.47\textwidth]{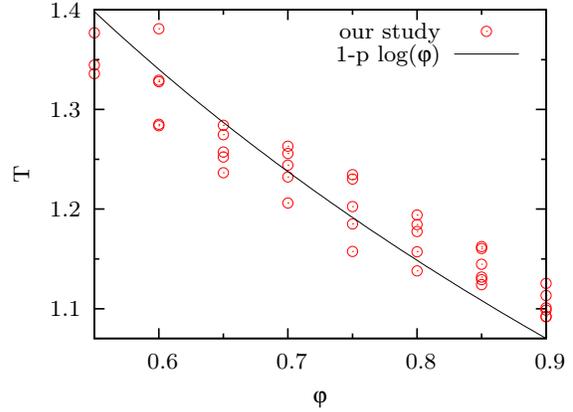}
\caption{Tortuosity as a function of porosity in the
two-dimensional model of overlapping circles computed using Eq.\ (\ref{tformula}). Symbols
are our numerical results, the solid line is the best fit to Eq.\
(\protect\ref{plog}). \label{tphi2d}}
\end{figure}

In our 
previous studies \cite{Matyka08,Koza09} we considered a similar model in which 
the porous matrix was modeled by overlapping squares and the tortuosity
was computed directly as an average over streamline lengths. We found 
that in that model the relation between tortuosity and porosity 
could be approximated by Comiti's and  Renaud's logarithmic  formula  \cite{Comiti89}
\begin{equation}
  \label{plog}
   T = 1 - p \ln \varphi
\end{equation}
with a fitting parameter $p=0.77$. 
The dependency of the tortuosity on the porosity in the present model of overlapping circles
is shown in Fig.~\ref{tphi2d}.
We fitted the data to Eq.\ (\ref{plog}) and found a good agreement for $p\approx 0.67$.

\subsection{Three-dimensional overlapping spheres}

As the final test of Eqs. (\ref{tformula}) and (\ref{tformula-sums}) we used them 
to find the flow tortuosity in a 
three-dimensional (3D) model of overlapping spheres. The
geometry was constructed similarly to the two-dimensional case described above. 
A regular grid of $90\times 90\times 90$ nodes was generated and freely overlapping spheres 
of radius 10 l.u. were deposited in it 
to reach the desired porosity value.
The system was assumed periodic in
the $x$ direction and no-slip walls were imposed at the four remaining walls. We ran
the LBM simulation for 10~000 time steps to reach the steady state and
then computed the tortuosity from Eq.~(\ref{tformula-sums}).

\begin{figure}
\centering
\includegraphics[width=.47\textwidth]{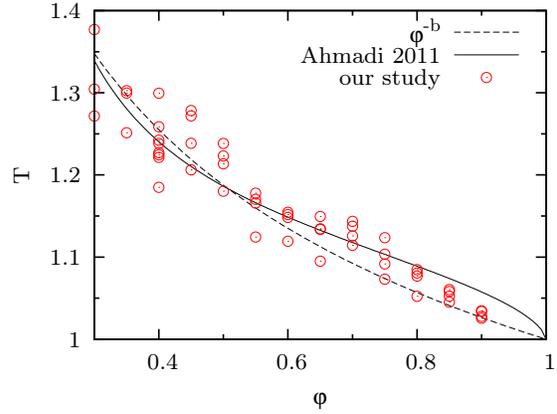}
\caption{
 Tortuosity, Eq.~(\protect\ref{tformula}), as a function of porosity in a three-dimensional model of overlapping spheres.
 Open symbols are our numerical results, the solid line is the best
 fit to Eq.~(\protect\ref{eq:ahmadi}), and 
  the dashed line is the best fit to $T=\varphi^{-b}$ with $b\approx 0.25$.
 \label{tortfig3d}
}
\end{figure}

The tortuosity-porosity dependence in our 3D model
is shown in Fig.~\ref{tortfig3d} (circles).
The solid line in this figure depicts a formula that was recently derived 
analytically for a very similar model  \cite{Ahmadi11,Hayati12}, 
\begin{equation}
 \label{eq:ahmadi}
   T = \sqrt{\frac{2\varphi}{3\left[1-B(1-\varphi)^{2/3}\right]}+\frac{1}{3}},
\end{equation}
where $B$ is a constant that depends on the shape of the obstacles and the lattice used.
The main difference between our model and that studied in \cite{Ahmadi11,Hayati12}
is that we distribute the spheres at random positions and allow them to overlap, 
whereas Eq.~(\ref{eq:ahmadi}) was derived for regular arrangements of impermeable, 
non-overlapping  objects of essentially arbitrary shape. 
Assuming that Eq.~(\ref{eq:ahmadi}) can be also used for non-regular arrangements of spheres, 
we fitted our data to this formula
using $B$ as a  free parameter.  This gave $B\approx 1.09$, which is a bit smaller than  
$B\approx1.209$ derived in \cite{Ahmadi11} for the cubic packing of spheres.
As seen in Fig.~\ref{tortfig3d}, our results are in good agreement with 
Eq.~(\ref{eq:ahmadi}) in the whole range of porosities.

Archie's law \cite{Archie42,Clennell97}
\begin{equation}
 \label{eq:Archie}
   T_\mathrm{el} = \varphi^{-n},
\end{equation}
is another formula for the tortuosity-porosity relation, often used for retardation tortuosities, 
especially the electrical and diffusional one. Together with Eqs.~(\ref{eq:tortuosity}), (\ref{eq:tortuosity-el}), 
Archie's law suggests $T = \varphi^{-b}$ with $b = (n-1)/2$. 
Our data can be fitted to this formula with $b\approx0.25$, see Fig.~\ref{tortfig3d},
which yields $n \approx 1.5$, which lies in  the range $1.3 \le n \le 3 $ reported in \cite{Clennell97}.
A more detailed analysis, involving a bigger number of larger systems, is necessary to verify which of the 
equations should be applied to this model, and  
whether Eq.~(\ref{eq:ahmadi}) actually applies to non-regular arrangements of obstacles.  

\section{Discussion and outlook}

We have presented several applications of our recently introduced 
method of calculating the tortuosity defined as a measure of the average elongation of 
fluid streamlines \cite{Duda11}. 
Starting from a simple model of an inclined channel, through a U-shaped channel, 
and ending at complex 2D and 3D geometries, we found a very good agreement of this 
method with other methods serving the same purpose. 
However, the main advantage of our method is that it does not require to find any streamlines,
which is a complicated, time-consuming and error-prone task, especially in realistic 3D geometries.
Instead, it allows to calculate the streamline tortuosity directly from the velocity field. 
Not only does this simplify numerical studies of this quantity, but  
should also greatly simplify experimental measurements of $T$.

To summarize, there are  several advantages of calculating the streamline tortuosity 
using the method of Ref. \cite{Duda11}. 
First, the formula is simple and flexible, allowing to compute the tortuosity
of practically any hydrodynamical fluid flow system in which the velocity field can be determined, 
whether numerically, analytically or experimentally. 
Second, this method solves the problem of the very existence of tortuosity 
as an average elongation of fluid paths. As a consequence, 
one can concentrate on the physical significance of this quantity, 
including finding its relation with numerous ``tortuosities'' defined as transport retardation factors.
In this context it is interesting to notice that since the streamline tortuosity 
can be expressed as the ratio of the average fluid velocity magnitude 
to the average fluid velocity along the macroscopic flow direction, 
it turns out to be closely related  to one of the most fundamental 
physical phenomena---momentum transfer. 
Third, this formula can be applied to other forms of transport in porous media, e.g. 
to diffusion or electric current \cite{Duda11}.  
Fourth, our formula can be used to flows in the fractal-like
\cite{Andrade07} or ramified structures \cite{Andrade98,Almeida99}, sytems which are not considered porous. 
For example,
one could compute hemodynamical tortuosity in the flow through
human artery system, in which a fundamental difference between the geometrical and
hydraulic tortuosities 
may be of profound importance for medical
diagnosis of arterial diseases \cite{Wolf01}.

\begin{theacknowledgments}
This work was supported by MNiSW (Ministry of Science and Higher Education)
Grant No. N N519 437939.
\end{theacknowledgments}

\bibliographystyle{aipproc}   % if natbib is available

\bibliography{tort}

\end{document}